\documentclass[aps,pra,twocolumn,superscriptaddress]{revtex4-1}
\usepackage{graphicx}
\usepackage{amsmath}
\usepackage{subfigure}
\usepackage{braket}
\begin{document}


\title{Casimir-Polder effect with thermally excited surfaces}


\author{A. Laliotis}
\email{athanasios.laliotis@univ-paris13.fr}
\affiliation{Laboratoire de Physique des Lasers, Universit{\'e} Paris 13, Sorbonne Paris-Cit{\'e}, F-93430, Villetaneuse, France}
\affiliation{CNRS, UMR 7538, LPL, 99 Avenue J.-B. Cl{\'e}ment, F-93430 Villetaneuse, France}
\author{M. Ducloy}
\affiliation{Laboratoire de Physique des Lasers, Universit{\'e} Paris 13, Sorbonne Paris-Cit{\'e}, F-93430, Villetaneuse, France}
\affiliation{CNRS, UMR 7538, LPL, 99 Avenue J.-B. Cl{\'e}ment, F-93430 Villetaneuse, France}
\affiliation{Departement of Physics and Applied Physics, School of Physical and Mathematical Sciences, Nanyang Technological University, 637371 Singapore}



\date{\today}

\begin{abstract}
We take a closer look at the fundamental Casimir-Polder interaction between quantum particles and dispersive dielectric surfaces with surface polariton or plasmon resonances. Linear response theory shows that in the near field, van der Waals, regime the free energy shift of a particle contains a thermal component that depends exclusively on the population/excitation of the evanescent surface polariton/plasmon modes. Our work makes evident the link between particle surface interaction and near field thermal emission and demonstrates how this can be used to engineer Casimir-Polder forces. We also examine how the exotic effects of surface waves are washed out as the distance from the surface increases. In the case of molecules or excited state atoms, far field approximations result in a classical dipole-dipole interaction which depends on the surface reflectivity and the mean number of photons at the frequency of the atomic/molecular transition. Finally we present numerical results for the CP interaction between Cs atoms and various dielectric surfaces with a single polariton resonance and discuss the implications of temperature and retardation effects for specific spectroscopic experiments.


\end{abstract}


\maketitle
The Casimir-Polder (CP) interaction between a polarisable quantum object (atom or molecule) and a surface arises from quantum fluctuations in vacuum. It's an excellent candidate for fundamental tests of cavity quantum electrodynamics and crucial for any experiments attempting to measure non-Newtonian gravity interactions \cite{harberpra2005, wolfpra2007}. CP forces are also relevant in physical chemistry playing an important role in the interpretation of physical phenomena such as atomic adsorption and desorption from hot surfaces or even surface chemistry and catalysis . The continuous urge for miniaturisation  has led to integrated devices, such as atom and molecule chips \cite{LinPRL2004, meekscience2009, pollockNJP2011, nshiinaturenanotech2013}, used for a variety of applications and more recently tapered nano-fibers were used to trap atoms at distances as small as 200 nm away from the surface \cite{gobanPRL2012, vetschprl2010, saguprl2007}, where atom-surface forces become exceedingly relevant. Novel trapping schemes that exploit the complexity of the van der Waals (vdW) potential of excited atoms have also been proposed \cite{changnatcommun2014}.  

The most basic description of the CP effect is that of a classical dipole interacting with its surface induced image. This approach is mostly valid in the vdW ($ z^{-3} $ law) regime, but QED theory \cite{casimirphysrev1948} revealed that when distances are larger than the wavelength corresponding to atomic transition, retardation effects scramble the interaction giving a $ z^{-4} $ distance dependence. For excited state atoms or molecules an additional contribution \cite{WSpra1984,WSpra1985} resembling the interaction of an antenna with its own reflected field has to be considered \cite{hindspra1991}. Thermal corrections to the CP force are analogous to the black body radiation induced corrections to the well known Lamb shift  \cite{hallblackbody}. In thermal equilibrium the problem has been considered by several authors \cite{bartonPRSLA1997, gorzaEPJD2006, EBSprl2010}. A novel behaviour was predicted when the surface and the vacuum are at different temperatures \cite{antezzaprl2005}. 

Accurate experimental demonstrations of the vdW law were given in a series of experiments performed with beams of Rydberg atoms \cite{hindsprl1992} as well as spectroscopic selective reflection experiments \cite{oriaepl1991, chevrollieroptlett1991} . Retardation effects were also demonstrated \cite{hindscp1993} with ground state sodium beam. Several experiments with cold atomic clouds have also been performed \cite{landraginpr1996, mohapatraepl2006, benderprl2010, benderprx2014} and a BEC positioned $6-12\mu m$ away from a silica surface was used to demonstrate the temperature dependence of the atom surface  interaction out of thermal equilibrium \cite{obrechtprl2007}.

In thermal equilibrium, temperature effects had remained elusive and were only very recently demonstrated using spectroscopic measurements in thermal vapour cells \cite{passeratlaserphysics2014, laliotisnatcommun2014} that probe atoms at distances on the order of 100 nm away from the surface. Critical to this experiment is the probing of excited state atoms that, much like molecules and in contrast to ground state atoms, have the advantage of presenting numerous dipole couplings in the mid and far infra-red. At these frequencies dielectrics support surface polariton modes whose thermal excitation creates  nearly monochromatic electromagnetic fields (compared to the well-known black body radiation) that evanescently decay away from the surface \cite{shchegrovprl2000}. 

Here we use quantum mechanical linear response theory to calculate the thermal CP interaction at all distances away the surface. By resumming the Matsubara frequency expansion we derive analytical expressions  which demonstrate that the thermal component of the vdW interaction  can be considered as a shift of the atomic levels due to near field thermal emission of surface modes. As such, its sign and strength depend exclusively on the relative position (detuning) of the atomic transitions compared to the frequency of the surface-polariton resonance. We show that the thermal excitation of surface polariton /plasmon modes can have observable effects even for low lying excited state or ground state atoms or molecules. We also examine scenarios where temperature changes can lead to a complete cancellation or change of sign of the atom-surface interaction. This allows engineering of CP forces with the use of temperature. We also derive analytical expressions for the resonant contribution to the CP interaction in the far field regime. The resonant contribution, dominant in many cases of interest (e.g. excited atoms and molecules) resembles a classical dipole-dipole interaction and depends on the surface reflectivity. Unlike the non-resonant CP predicted behaviour for ground state atoms \cite{taillandierpra2014}, the resonant contribution preserves a strong anisotropic component, characteristic of dipole-dipole interactions. We finally discuss the transition between the two regimes using numerical calculations. We demonstrate that controlling the CP interaction with temperature is no longer possible as the distance from the surface increases. We also show that retardation effects are relevant even for spectroscopic experiments at nanometric distances away from the surface. 

\section{Casimir-Polder interaction}

We start by considering the  CP free energy shift of a quantum particle $\Delta F_{a}$ at a given energy level $ \ket{a} $, which is the sum of a resonant  and a non-resonant contribution \cite{WSpra1984, WSpra1985}. We follow the formalism of M-P Gorza et. al. \cite{gorzaEPJD2006}, describing the free energy shift at a finite temperature T. 
\begin{equation}
\Delta F_{a}=\Delta F_{a}^{r}+\Delta F_{a}^{nr}
\label{eqn1}
\end{equation}
To simplify the notation we will expand our reasoning for a two level $ \ket{a},  \ket{b} $ system. For a real multilevel system one simply has to sum all the contributions of all individual dipole couplings. 

At a finite temperature the non resonant term is given by the following sum :  
\begin{equation}
\Delta F_{a}^{nr} =-2\frac{k_{B} T}{\hbar}  {\sum_{k=0}^{\infty}}^{\prime} \mu_{\alpha}^{ab}\mu_{\beta}^{ba}G_{\alpha\beta}(z,i\xi_{k}) \frac{\omega_{o}}{\xi_{k}^2+\omega_{o}^{2}}
\label{eqn2}
\end{equation}
We use the Einstein notation , implying a summation over the index variables $\alpha$ and $\beta$ that denote the Cartesian coordinate components. The prime symbol signifies that the first term of the sum should be multiplied by 1/2. The transition frequency $ \omega_{o}=(E_{b}-E_{a})/\hbar$ depends on the energy difference between the two levels. It takes positive or negative signs depending on the nature of the coupling (absorption or emission). Also  $\xi_{k}=2 \pi \frac{k_{B} T}{\hbar} k$ are the Matsubara frequencies, $ \mu_{\alpha}^{ab}$ and $\mu_{\beta}^{ba}$ are the dipole moment matrix elements and $G_{\alpha\beta}(z,i\xi_{k})$ are the components of the linear susceptibility matrix of the reflected field, defined in \cite{WSpra1984, WSpra1985}. In the general case, the linear susceptibility gives the reflected displacement field at a point $ \vec{r}$ due to a dipole  $\vec{\mu}(\omega)$, oscillating at a frequency $\omega$, positioned at $ \vec{r^{\prime}}$, via the relation $\vec{D}( \vec{r}, \vec{r^{\prime}},\omega)=\stackrel{\leftrightarrow}{G}( \vec{r},\vec{r^{\prime}}, \omega) \vec{\mu}(\omega) $. In our case $\stackrel{\leftrightarrow}{G}$  is evaluated for $ \vec{r}=\vec{r^{\prime}}$, because we're interested in dipole interacting with their own reflected field. Due to the cylindrical symmetry $\stackrel{\leftrightarrow}{G}$ is only a function of frequency and the distance $ z $ of the dipole from the reflecting wall.

The resonant part of the CP shift is written as \cite{gorzaEPJD2006}:
\begin{equation}
\Delta F_{a}^{r} =n(\omega_{o},T) \mu_{\alpha}^{ab}\mu_{\beta}^{ba} Re\left[ G_{\alpha\beta}(z,|\omega_{o}|)\right]
\label{eqn3}
\end{equation} 
The mean occupation number $n(\omega_{o},T)=\frac{1}{e^{\frac{\hbar\omega_{o}}{k_{B}T}}-1}$ of a mode according to Bose-Einstein statistics is here extended to negative frequencies. Note that in the case of virtual emission $\omega_{o}<0$ the sign changes and an additional photon due to spontaneous emission is added, $n(\omega_{o},T)=-\left[ 1+n(-\omega_{o},T) \right] $.

The problem reduces to calculating the linear susceptibility function. We will focus on the simple geometry of a semi-infinite surface with a bulk dielectric constant $\epsilon(\omega)$ and an atom that is in the vacuum . The diagonal terms of the linear susceptibility \citep{WSpra1984, WSpra1985, gorzaunpublished} calculated for an imaginary frequency $i\xi$ are :
\begin{equation}
G_{xx}(z,i\xi)=G_{yy}(z,i\xi)=\frac{\xi^3}{2c^3 } \int_{1}^{\infty}e^{-\frac{2 \xi z u}{c}}\left( u^2 R^p-R^s \right )du 
\label{eqn4}
\end{equation}
\begin{equation}
G_{zz}(z,i\xi)=\frac{\xi^3}{c^3 } \int_{1}^{\infty}e^{-\frac{2 \xi z u}{c}}\left( u^2 -1 \right )R^p du
\label{eqn5}
\end{equation}
In the above equations $R^p$ and $R^s$ are the Fresnel reflection coefficients and $u$ is a dummy integration variable. The problem has no simple analytical solutions apart from the famous case of an ideal conductor or an ideal dispersion-less dielectric. For a real surface one has to resort to  numerical simulations except in the limiting cases when $\frac{2 \xi z}{c}\gg 1$, i.e the long range case where retardation effects are important, or $\frac{2 \xi z}{c} \ll 1$ which is the electrostatic or van der Waals regime.

\section{van der Waals interaction}

Since the first experimental demonstration of retardation effects \cite{hindsprl1992} the vdW interaction has been mostly considered as an electrostatic limit of the CP interaction. A QED description is however necessary when the atom is in the presence of a hot surface. Here we pursue further the results of  \cite{gorzaEPJD2006} by considering real dielectrics with one polariton resonace in order to illuminate the physics of CP interactions in the presence of polaritons or plasmons. We show that excited surface waves create intense thermal fields in the vicinity of the surface (near field thermal emission) that offer a way to control the vdW interaction, in a way that is not possible in the far-field regime.

In the near field, vdW regime ($ z \ll \frac{\lambda_{o}}{4 \pi}$, where $\lambda_{o}$ is the atomic transition wavelength) the linear susceptibility has an analytical solution given by \cite{WSpra1985}:
\begin{equation}
G_{xx}(z,\omega)=G_{yy}(z,\omega)=\frac{1}{(2z)^3} S(\omega)
\label{eqn6}
\end{equation}
\begin{equation}
G_{zz}(z,\omega)=\frac{2}{(2z)^3} S(\omega)
\label{eqn7}
\end{equation} 
Eqns. (6,7) are valid for both real and imaginary frequencies when $\frac{2 |\omega| z}{c} \ll 1$. Here $S$ is the frequency dependent image coefficient which is a function of the bulk dielectric constant $\epsilon(\omega)$. It is given by:
\begin{equation}
S(\omega)=\frac{\epsilon(\omega)-1}{\epsilon(\omega)+1} 
\label{eqn8}
\end{equation}

We start by using a single resonance model to describe the dielectric constant of the surface. 
\begin{equation}
\epsilon \left( \omega \right )=\epsilon_{inf}+\frac{(\epsilon_{st}-\epsilon_{inf})\omega_{T}^{2}}{\omega_{T}^{2}-\omega^{2}-i \Gamma \omega}
\label{eqn9}
\end{equation}
where $\epsilon_{inf}$ and $\epsilon_{st}$ are constants giving the dielectric constant at the two extreme ends of the spectrum, $\Gamma$ is phenomenological decay rate and $\omega_{T}$ is the transverse frequency of oscillations. The above equation models dielectrics with one surface polariton resonance. The validity of eqn. (9) is limited to a certain frequency range (see also discussion in \citep{bartonPRSLA1997}), however, it accurately describes the CP interaction between many atom-surface systems.  A more realistic model should account for multiple resonances, however this scenario will not be considered here. 

We first assume that $\Gamma=0$. In this case, the surface polariton frequency is given by $\omega_{S}=\omega_{T}\sqrt{\frac{\epsilon_{st}+1}{\epsilon_{inf}+1}} $ and the image coefficient becomes:
\begin{equation}
S \left( \omega \right )=S_{inf}+(S_{st}-S_{inf})\frac{\omega_{S}^{2}}{\omega_{S}^{2}-\omega^{2}}
\label{eqn10}
\end{equation}
Here $S_{inf} $ and $S_{st} $ represent the values of $S(\omega)$ calculated for $\epsilon_{inf}$ and $\epsilon_{st}$ respectively.  Eqn.(\ref{eqn10}) diverges at the polariton frequency but this is a small price to pay for keeping our analytical expressions simple with a clear physical interpretation. 

Reporting the linear susceptibility eqns. (\ref{eqn6}, \ref{eqn7}) and the dielectric constant eqn. (\ref{eqn9}) into eqn. (\ref{eqn2}), we can sum the Matsubara frequency expansion and arrive to the following analytical expression for the non-resonant free energy shift of the vdW interaction:
\begin{equation}\begin{split}
\Delta F_{a}^{nr}=-\frac{C_{3}^{pc}}{z^3} [ Re \left[ S(\omega_{o}) \right] \coth \left( \frac{\hbar \omega_{o}}{2k_{B}T} \right) + \\
\frac{\left( S_{st} -S_{inf} \right) \omega_{o} \omega_{S} }{\omega_{o}^{2}-\omega_{S}^{2}}  \coth \left( \frac{\hbar \omega_{S}}{2k_{B}T} \right) ]
\label{eqn11}
\end{split}\end{equation}

Here the constant $C_{3}^{pc}=\frac{|\bra{a}\mu \ket{b}|^2}{12}$ represents the vdW coefficient for a perfect conductor. We have also used the relation $\coth \left( \frac{\hbar \omega}{2k_{B}T} \right)=2n(\omega,T)+1 $. One notices the characteristic $z^{-3}$ dependence of the vdW interaction. The terms inside the brackets in eqn. (\ref{eqn11}) depend only on the dielectric properties of the surface and on temperature. We refer to them as the non-resonant part of the image coefficient. The seemingly complicated temperature dependence of the non-resonant term, given by a complex sum over imaginary frequencies, essentially reduces down to the number of photons in the atomic and polariton frequencies. On the other hand the resonant part of the shift, eqn. (\ref{eqn3}), only depends on the number of photons in the atomic frequency and is given by:
\begin{equation}
\Delta F_{a}^{r}=-\frac{C_{3}^{pc}}{z^3} \left[  Re \left[ S(\omega_{o}) \right]-Re \left[ S(\omega_{o}) \right] \coth \left( \frac{\hbar \omega_{o}}{2k_{B}T} \right) \right]
\label{eqn12}
\end{equation}

We refer to the term inside the brackets in eqn. (\ref{eqn12}) as the resonant part of the image coefficient. A simple inspection of the above equations reveals that the temperature dependence of the resonant contributions cancels out with the first term of the non-resonant contribution. Adding eqn. (\ref{eqn11}) and eqn. (\ref{eqn12}) we find the total free energy shift $\Delta F_{a}=-\frac{C_{3}^{pc}}{z^3}r\left(\omega_{o},T \right)=-\frac{C_{3}}{z^3}$. Here, $C_{3}$ is the vdW coefficient and $r\left(\omega_{o},T \right)$ is the image coefficient given by:

\begin{equation}
r\left(\omega_{o},T \right)=   Re \left[ S(\omega_{o})\right]+ 
\frac{\left( S_{st} -S_{inf} \right) \omega_{o} \omega_{S} }{\omega_{o}^{2}-\omega_{S}^{2}}   \coth \left( \frac{\hbar \omega_{S}}{2k_{B}T} \right) 
\label{eqn13}
\end{equation}

The above equation gives the temperature dependence of the vdW interaction. It depends strictly on thermal fluctuations at the polariton frequency i.e on the thermal excitation of the evanescent surface polariton modes and not on the number of thermal photons at the atomic frequency. In the two extremes of the spectrum, when $\omega_{o}\gg\omega_{S}$ and $\omega_{o}\ll\omega_{S}$ all temperature dependence vanishes, in accordance with previous results given for an perfect conductor \cite{EBSprl2010}. In consistence with the classical picture, when material dispersion is neglected the vdW attraction is independent of temperature. Similar results can be obtained with the methodology of \cite{bartonPRSLA1997} but eqn. (13) is more general since it is valid for all temperatures, including T=0 and includes both virtual absorption ($\omega_{o}>0$) and emission ($\omega_{o}<0$). In the limit of high photon number the temperature dependent part of eqn. (13) can also be derived by calculating the Stark shift induced by the thermally populated evanescent surface polariton  modes whose density of states, $\rho_{D}$, is given by \cite{shchegrovprl2000, jonesPSS2013}:
\begin{equation}
\rho_{D} ( \omega, T, z)= \frac{1}{8 \pi^2 \omega z^3} Im \left[ \frac{\epsilon-1}{\epsilon+1} \right]
\label{eqn14}
\end{equation}
This last approach is analogous to the one used in \cite{farleyPRA1981} in order to calculate the atomic shifts due to black body radiation. 

The calculation becomes significantly more cumbersome if $\Gamma \neq 0 $, but analytical expressions can also be found. In most cases of practical interest ($ \Gamma \ll \omega_{S}$) many terms can be neglected and the image coefficient becomes:
\begin{equation}\begin{split}
r\left(\omega_{o},T \right)&=   Re \left[ S(\omega_{o})\right]+ \\
  &\frac{2\omega_{o} \omega_{S} \left( S_{st} -S_{inf} \right)\left[ \Gamma^2+2(\omega_{o}^2-\omega_{S}^2) \right] }{\left[ \Gamma^2+2(\omega_{o}^{2}-\omega_{S}^{2})\right]^2 + 4\Gamma^2 \omega_{S}^2}  \coth \left( \frac{\hbar \omega_{S}}{2k_{B}T} \right) 
\label{eqn15}
\end{split}\end{equation}
Within the limits of eqn. (\ref{eqn15}) the cancellations that make the vdW interaction independent of the number of photons at the transition frequency $\omega_{o}$ are still valid. The above approximation deviates from the exact solution in the vicinity of the polariton resonance only by a few percent, whereas in the rest of the spectrum differences are negligible.

\begin{figure}
\includegraphics[width=90mm]{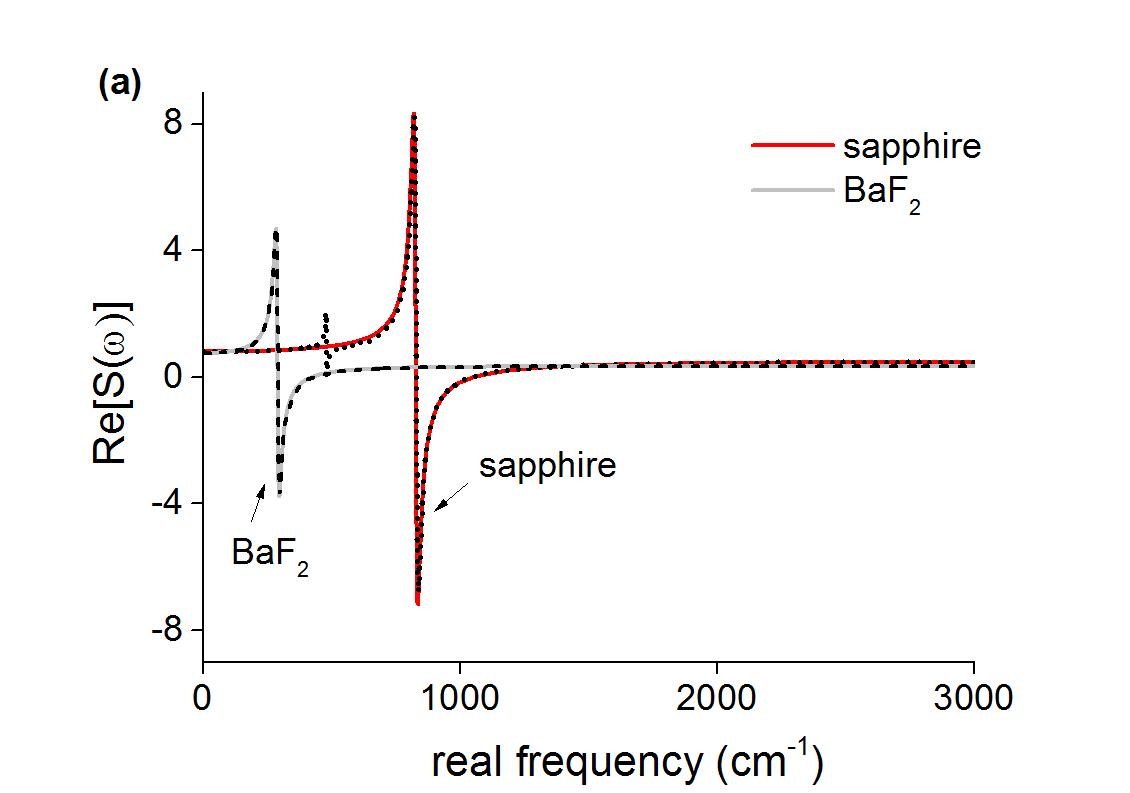}%

\includegraphics[width=90mm]{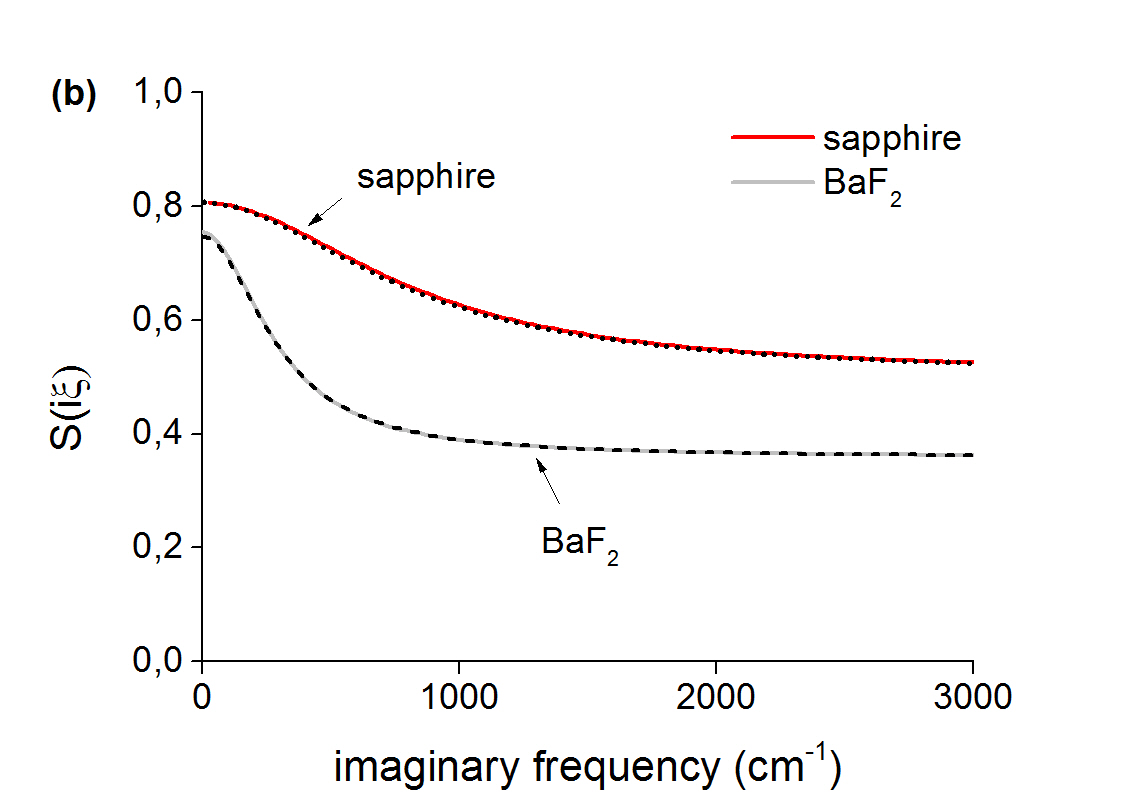}%
\caption{(Color online)(a) The real part of the surface response for real frequencies. The results of the hydrodynamic model,eqn. (\ref{eqn9}) are shown for sapphire (red or dark grey) and $\textrm{BaF}_{2}$ (grey). The black dotted (sapphire) and dashed ($\textrm{BaF}_{2}$) lines represent the results of the elaborate model of ref.\cite{passeratJP2009}. (b) The surface response for imaginary frequencies using the exact same colour/line coding. The dielectric constant and therefore the surface response are strictly real in this case.   \label{Fig1}}
\end{figure}

To illustrate the importance of these results we start by giving in Table I a list of all the parameters involved in eqn. (\ref{eqn9}) i.e $\epsilon_{st}$, $\epsilon_{inf}$, $\omega_{T}$ and $\Gamma$ for some dielectrics. Amongst them, sapphire is very commonly used in atom-surface interaction experiments whereas $\textrm{CaF}_{2}$ and $\textrm{BaF}_{2}$ have been considered for this purpose (see \cite{passeratJP2009, passeratlaserphysics2014}) due to their isolated surface resonances in the relatively far infra-red.   $\textrm{SiC}$ is most commonly used for near field thermal emission measurements \cite{GreffetNature2002, BabutyPRL2013}. Here we explore the potential interest of performing CP measurements  with this material. We also restrict ourselves to dielectrics whose dielectric constant is adequately described by the hydrodynamic model throughout the visible, near and far infra-red frequency range. In the case of $\textrm{CaF}_{2}$, $\textrm{BaF}_{2}$ and sapphire the parameters are deduced by fitting eqns. (\ref{eqn8},\ref{eqn9}) to the experimental data given in ref. \cite{passeratJP2009}, whereas for SiC the parameters are taken directly from \cite{shchegrovprl2000}. In Fig.\ref{Fig1}(a) we plot the real part of the surface response vs real frequencies ($\omega/2\pi$) for both sapphire and $\textrm{BaF}_{2}$. The results of eqn. (\ref{eqn8}) (coloured lines) are compared to those of the a more elaborate model described in \cite{passeratJP2009} (black lines). The surface response for imaginary frequencies ($\xi/2\pi$) is shown in Fig.\ref{Fig1}(b). The hydrodynamic model of eqn. (\ref{eqn9}) reproduces very well the surface response of the above dielectrics. The only observable discrepancy between the two models is due to the existence of an additional, albeit much smaller, surface resonance at $480 \ cm^{-1}$, in the case of sapphire. One should not be deceived into thinking that eqn. (\ref{eqn9}) is a perfect model of the dielectric constant itself. However, Fig.\ref{Fig1} clearly demonstrates that our simple model can be reliably used to predict the van der Waals free energy shifts of atoms/molecules against any of the dielectric surfaces of Table I.

\begin{table}[ht]
\caption{Parameters used to model the dielectric constant of sapphire, $\textrm{CaF}_{2}$, $\textrm{BaF}_{2}$ and $\textrm{SiC}$.}  
\centering 
\begin{tabular}{c c c c c c} 
\hline\hline 
 & $\epsilon_{inf} $ & $\epsilon_{st}$ & $\frac{\omega_{S}}{2 \pi} $ ($cm^{-1}$) & $\frac{\omega_{T}}{2 \pi} $ ($cm^{-1}$)&  $\frac{2 \pi \Gamma}{\omega_{S}}$ \\ [0.5ex] 
\hline 
$\textrm{BaF}_{2}$  & 2.12 & 7.16 &  291 & 179.9 & 0.047\\ 
$\textrm{CaF}_{2}$ & 2.02 & 6.82 & 416.2 & 258.6 & 0.063\\
Sapphire  & 3.03 & 9.32 & 828.9 & 518 & 0.02\\
SiC & 6.7 & 10 & 947.8 & 793 & 0.005 \\ [1ex] 
\hline 

\end{tabular}
\label{table:nonlin} 
\end{table}

\begin{figure}
\includegraphics[width=90mm]{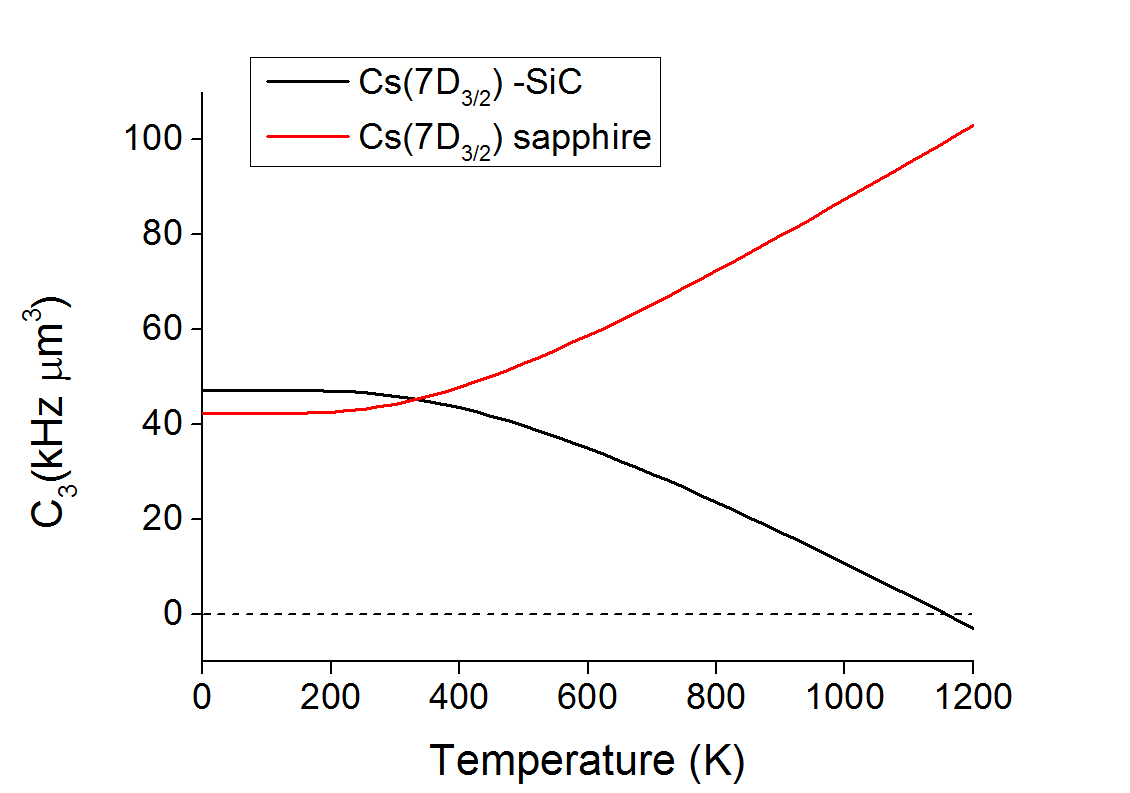}%
\caption{(Color online) The vdW coefficient as a function of temperature for a Cs($7D_{3/2}$) atom against a sapphire (red or dark grey) and a SiC (black) surface.  \label{Fig2}}
\end{figure}

Having established the validity and the main features of our model we can proceed to calculate the interaction between some realistic atom surface systems. In Fig. \ref{Fig2} we plot the $C_{3}$ coefficient as function of temperature for $Cs^{*}(7D_{3/2})$ against sapphire and SiC. The $C_{3}$ coefficient for this atomic level depends strongly on the $7D_{3/2} \rightarrow 5F_{5/2}$ dipole coupling at $923 cm^{-1}$. The sapphire resonance at $829 cm^{-1}$ is on the red side of this transition frequency leading to an increase of the vdW coefficient as a function of temperature as demonstrated experimentally in ref. \cite{laliotisnatcommun2014}. Conversely, the \textit{same atom}  near a SiC surface displays a completely different behaviour. The SiC resonance at $945 cm^{-1}$, is on the blue side of the transition frequency, thus the vdW coefficient now decreases with temperature. According to our theoretical estimates the fundamental long-range atom-surface interaction should be null at $T \sim 1200 K$, a temperature range which could be experimentally achievable. Fig. \ref{Fig2} shows the possibility of controlling the CP interactions close to dispersive surfaces via temperature. Other atom-surface systems have been considered for achieving repulsive vdW potentials \citep{passeratJP2009} at finite temperatures but so far there has been no experimental proof of this effect \citep{passeratlaserphysics2014}.

It is also of interest to estimate the effects of surface polaritons for ground state atoms. This situation is relevant for most experimental measurements using with cold atoms \cite{wolfpra2007, obrechtprl2007, landraginpr1996, benderprl2010}. The finite temperature corrections ($T=300K$) in the near field regime are in this case $0.02\%$, $0.16 \%$,$1.29\%$ and $1.75\%$ for SiC, sapphire ,$\textrm{CaF}_{2} $ and $\textrm{BaF}_{2} $ respectively. These numbers indicate that near field thermal emission has negligible effects compared to the experimental precision of most experiments so far, but it could have implications in the case of precision experiments aiming at putting new limits to the existence of Non-Newtonian gravity forces \cite{wolfpra2007, harberpra2005}. The above numbers should be considered as indicative because we have used the simplified model of eqn. (\ref{eqn9}) and they are calculated in the vdW electrostatic limit , which restricts their validity to very small distances away from the surface.

\section{Far field approximation}

The behaviour of the non-resonant term of eqn.(\ref{eqn1}) has been discussed extensively in the past (see e.g. \cite{EBSprl2010, obrechtprl2007, McLachlan1963} and references therein). We'll just remind here that at $T=0$ there's a passage from a $z^{-3}$ to a $z^{-4}$ law for the energy shift at large distances from the surface. At a finite temperature $T$ there's a second cross over to a $z^{-3}$ law when $z \gg \frac{1}{4 \pi}  \frac{\hbar c}{k_{B} T}$, where $\lambda_{T}= \frac{\hbar c}{k_{B} T}$ is usually referred to as the thermal wavelength. In this case the sum of eqn. (\ref{eqn2}) is dominated by the first term \cite{McLachlan1963} and therefore gives:

\begin{equation}
\Delta F_{a}^{nr}=-2\frac{\epsilon(0)-1}{\epsilon(0)+1} \frac{k_{B} T}{\hbar \omega_{o}}  \frac{C^{pc}_{3}}{z^{3}}
\label{eqn16}
\end{equation}

In the far field the free energy shift is independent of material dispersion. Only on the DC or 'static' value of the dielectric constant at zero frequency come into play. This far field or high temperature limit is sometimes refered to as the Lifshitz regime \cite{obrechtprl2007}.

The behaviour of the resonant term in the far field is rather different. Using the imaginary frequency formulas above we can see that in the far field the exponential decays rapidly. The multiplying functions $\left( u^2 R^p-R^s \right )$ and $\left( u^2 -1 \right ) R^p$  vary rather slowly and one can find approximate expressions for the linear susceptibility by Taylor expanding them around 1 to the lowest non-zero order. The real part of the linear susceptibility for real frequencies $\omega_{o}$ is the relevant quantity for the resonant term of the free energy shift, which is given by:

\begin{equation}
Re[G_{xx,yy}(z,\omega_{o})]=\frac{k_{o}^2}{2z_{o}}|\rho(\omega_{o})|cos(2k_{o} z_{o}+\phi(\omega_{o}))
\label{eqn17}
\end{equation} 
\begin{equation}
Re[G_{zz}(z,\omega_{o})]=-\frac{k_{o}}{2z_{o}^{2}}|\rho(\omega_{o})|sin(2k_{o} z_{o}+\phi(\omega_{o}))
\label{eqn18}
\end{equation}
where $\rho(\omega_{o})=|\rho(\omega_{o})| e^{i \phi(\omega_{o})}$ is the frequency dependent complex reflection coefficient of the surface at normal incidence. Eqns. (\ref{eqn17}, \ref{eqn18}) are valid when  $z \gg \frac{\lambda_{o}}{4 \pi}$, where $\lambda_{o}$ is the transition wavelength. As has been pointed out before \cite{hindspra1991} the resonant term resembles the interaction of an  antenna with its own reflected field, which oscillates between repulsion and attraction with a period of $\lambda_{o}/2$. In the case of a perfect conductor, eqns.(\ref{eqn17}, \ref{eqn18}) lead to the cavity QED shifts reported in \cite{hindspra1991}. The term is highly anisotropic since a dipole antenna does not radiate on its axis. For this reason $Re[G_{zz}(z,\omega_{o})]$ decays much faster than $Re[G_{xx,yy}(z,\omega_{o})]$. In reality this term will dominate the far field of CP interaction for excited state atoms or molecules \cite{EBSpra2009a, EBSpra2009b}. At very large temperatures or at very large distances from the surface it will eventually dominate the interaction between a surface and a ground state atom. 

\begin{figure}
\includegraphics[width=90mm]{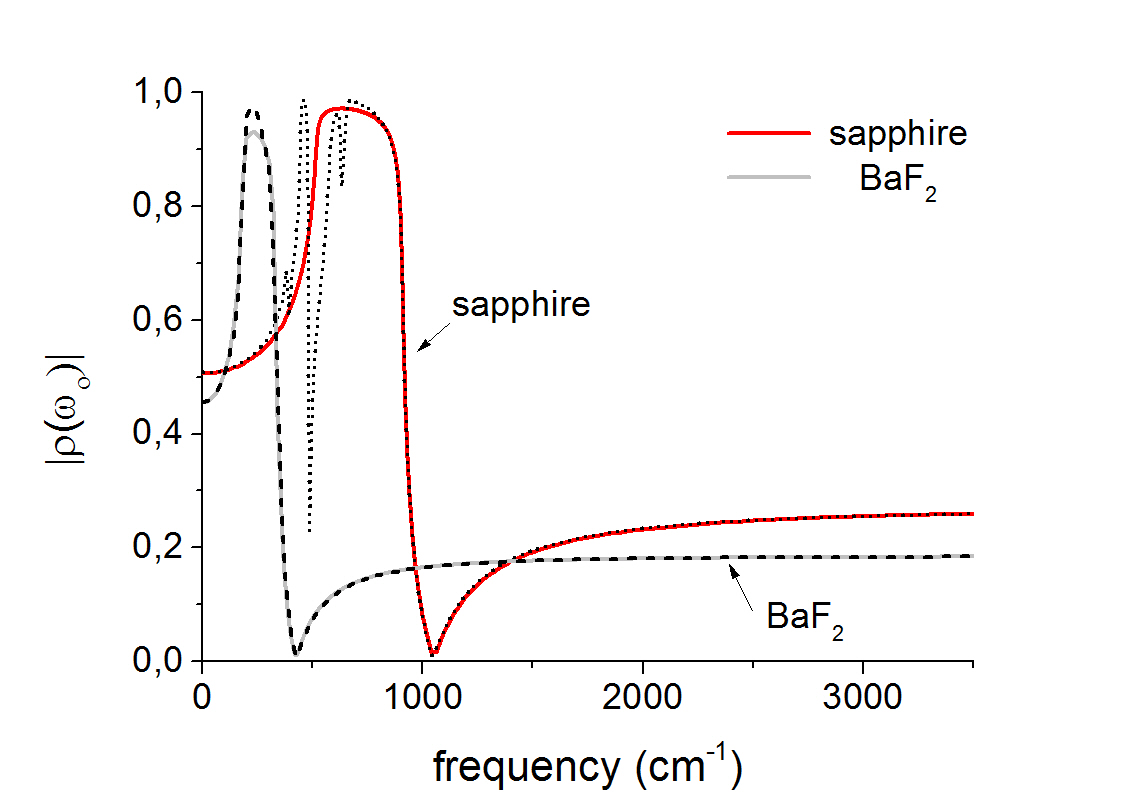}%
\caption{(Color online)The absolute value of reflection coefficient as a function of frequency. The solid lines represent the reflection coefficient as calculated using the hydronamic model of eqn. (\ref{eqn9}) in the case of sapphire (red or dark grey) and $\textrm{BaF}_{2} $ (grey). The black dotted (sapphire) and dashed ($\textrm{BaF}_{2} $) lines represent the results of the model described in ref. \citep{passeratJP2009}.   \label{Fig3}}
\end{figure}

In Fig.\ref{Fig3} we plot the reflection coefficient as a function of frequency. The limitations of the hydrodynamic model are here visible, especially for sapphire. The discrepancies are due to additional resonances that have been neglected \cite{barkerPR1963}. The most noteworthy feature of Fig.\ref{Fig4} is that the reflection coefficient does not display the same variations as the surface response which changes sign around the surface polariton frequency (Fig. \ref{Fig1}). This shows that exotic effects such as resonant vdW repulsion \cite{failacheprl1999} or temperature effects due to the thermal excitation of surface polaritons \cite{laliotisnatcommun2014} should be washed out as the distance form the surface increases.

\section{Discussion}

\begin{figure}
\includegraphics[width=90mm]{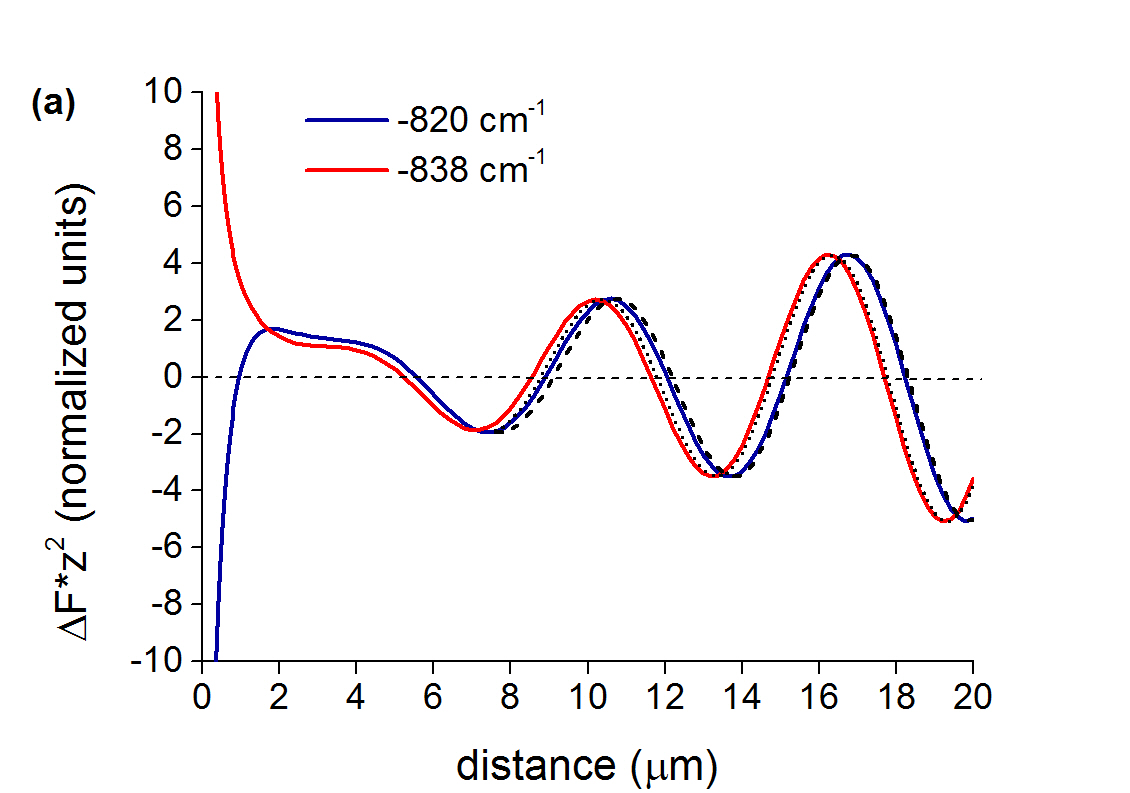}%

\includegraphics[width=90mm]{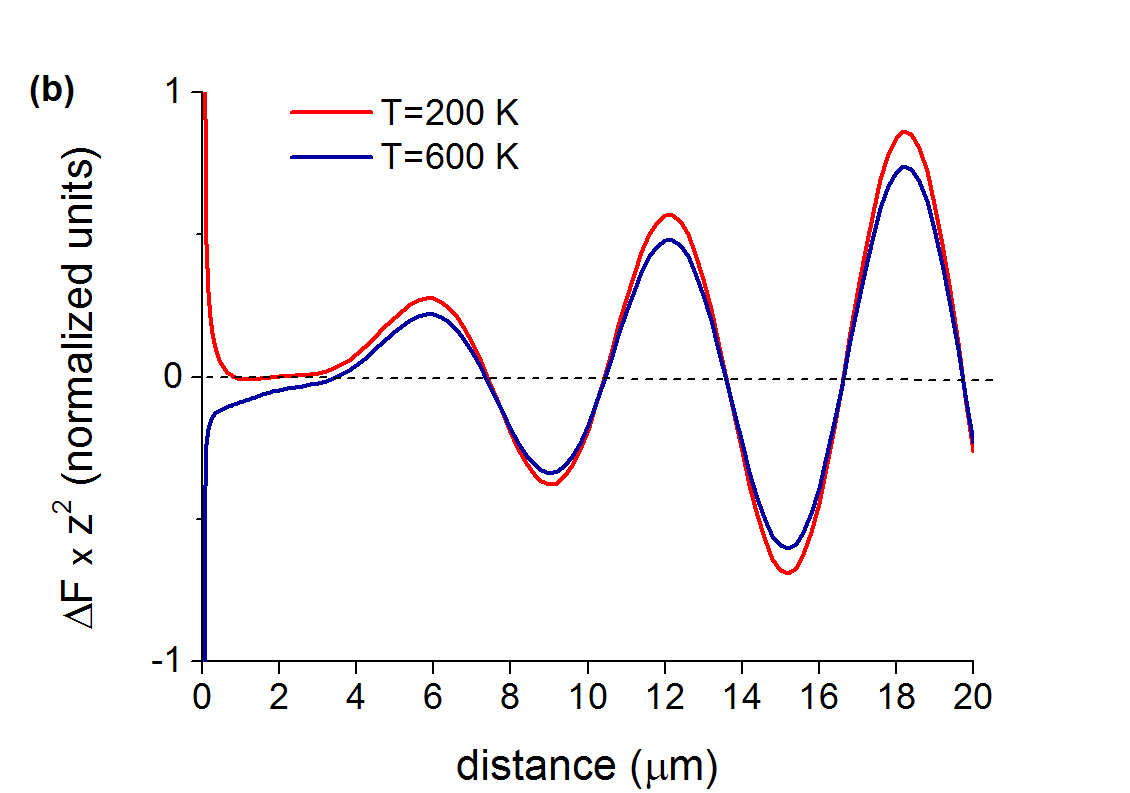}%
\caption{ (Color online)(a) The free energy shift multiplied by $z^{2}$ for $-820 cm^{-1}$ (blue or black) and $-838 cm^{-1}$ (red or dark grey) downward couplings against a sapphire surface at T=300 K. The dashed and dotted lines represent the free energy shift as given by the far field approximation of eqn (\ref{eqn17},\ref{eqn18}) (b)  The free energy shift multiplied by $z^{2}$ for a $-830 cm^{-1}$ downward coupling against a $\textrm{BaF}_{2} $ surface for $T=200K$ (blue or black) and $T=600K$ (red or dark grey) lines.\label{Fig4}}
\end{figure}

In order to demonstrate the effects of retardation and thermal excitation of the surface we focus our attention in some concrete examples. We first choose two  couplings (virtual emission) with frequencies $-820 cm^{-1}$ and $-838 cm^{-1}$ corresponding at the two extrema of the sapphire image coefficient, close to the sapphire resonance at $829 cm^{-1}$ (Fig.\ref{Fig1}). In Fig. \ref{Fig4}a we plot the normalised CP shift due to these couplings as a function of distance from a sapphire surface. The shift is calculated by numerically integrating eqns. (\ref{eqn4}, \ref{eqn5}) using the hydrodynamic model for sapphire's dielectric constant, without any approximations. In the near field the two dipole couplings present shifts of opposite sign. The first (blue line at $-820 cm^{-1}$) corresponds to a large attraction whereas the second (red line at $-838 cm^{-1}$) corresponds to a repulsion. This exotic effect persists only in the nanometric scale, at distances smaller than $1\mu m$. The two curves converge for larger distances, oscillating from attraction to repulsion in a very similar fashion. The reader should note that even though the amplitude of these oscillations is almost the same, there is a noticeable phase difference between them. The estimated free energy shifts using the far field approximation of eqns. (\ref{eqn17}, \ref{eqn18}) are also shown as black dashed and dotted lines. At large separations ($z > 10 \mu m $)they coincide almost perfectly with the numerical calculations. In Fig. \ref{Fig4}b we plot the free energy shift due downward coupling at $-830 cm^{-1}$ for a $\textrm{BaF}_{2} $ surface at two different temperatures $T=200 K$ and $T=600 K$. Due to the thermal excitation of the  $\textrm{BaF}_{2} $ surface polariton at $291 cm^{-1}$ the vdW interaction changes sign from attraction to repulsion. The near field temperature dependence is governed by the number of thermal photons at the the polariton frequency  $\omega_{S}$ of $\textrm{BaF}_{2}$ , as described by eqn.(\ref{eqn13}). Conversely the temperature dependence in the far field is related to the number of thermal photons at the transition frequency $\omega_{o}$, as can be seen by eqns. (\ref{eqn3},\ref{eqn17},\ref{eqn18}). Here again the effects of the evanescent polariton modes are only present in the nanometric range\cite{DorofeyevLP2013, jonesPSS2013 }. The transition frequencies used in Fig.\ref{Fig4} are in the vicinity of the $7\textrm{P} \rightarrow 6\textrm{D}$ transitions of both Cs and Rb. The vdW interaction of  $\textrm{Cs}^{*}(6\textrm{D}_{3/2})$, $\textrm{Rb}^{*}(6\textrm{D}_{3/2})$ and $\textrm{Rb}^{*}(6\textrm{D}_{5/2})$ was experimentally investigated in the past with selective reflection experiments in vapour cells \cite{failacheEPJD2003}.

\begin{table}[ht]
\caption{Individual contributions to the $C_{3}$ coefficient of the most important dipole couplings for Cs($6S_{1/2}$) and Cs($6P_{1/2}$) close to a perfect conductor. Transition probabilities are taken from ref. \cite{Lindgard1975} }  
\centering 
\begin{tabular}{c c c c} 
\hline 
\textbf{Cs($\textbf{6}\textrm{S}_{\textbf{1/2}}$)} &   $\lambda (\mu m) $  &  $\omega (cm^{-1}) $   &    $C_{3}^{pc} (kHz \mu m^3) $  \\ [0.5ex] 
\hline 
$6\textrm{P}_{1/2}$  & 0.894 & 11178.24 & 0.94 \\ 
$6\textrm{P}_{3/2}$ & 0.852 & 11732.35 & 1.62 \\
\hline
\\ [1ex] 
\end{tabular}

\begin{tabular}{c c c c} 
 \hline 
\textbf{Cs($\textbf{6}\textrm{P}_{\textbf{1/2}}$)}&   $\lambda (\mu m) $  &  $\omega (cm^{-1}) $   &    $C_{3}^{pc} (kHz \space \mu m^3) $  \\ [0.5ex] 
\hline 
$6\textrm{S}_{1/2}$  & -0.894 & -11178.24 & 0.94 \\ 
$5\textrm{D}_{3/2}$ & 3.01 & 3321.25 & 1.64 \\
$7\textrm{S}_{1/2}$ & 1.36 & 7357.27 & 0.59 \\
$6\textrm{D}_{3/2}$ & 0.876 & 11410.65 & 0.64 \\
$7\textrm{D}_{3/2}$ & 0.673 & 14869,62 & 0.15 \\ [1ex] 
\hline 

\end{tabular}
\label{table:2} 
\end{table}

\begin{figure}
\includegraphics[width=90mm]{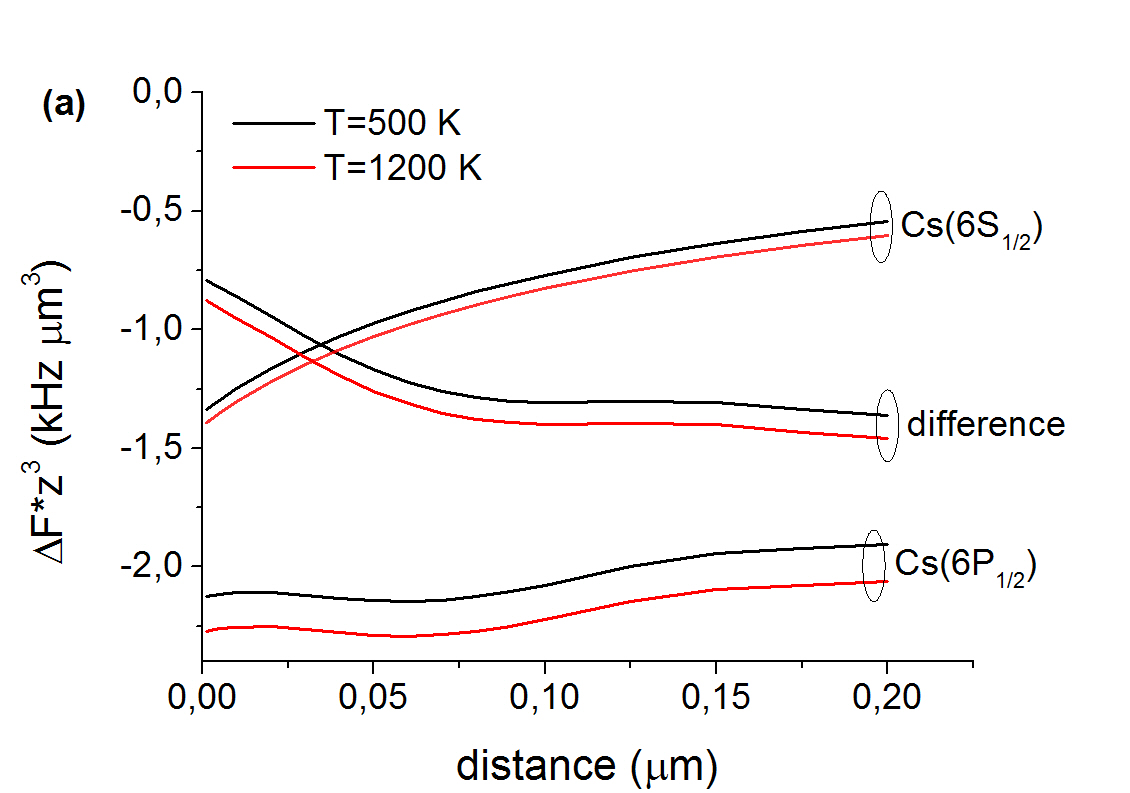}%

\includegraphics[width=90mm]{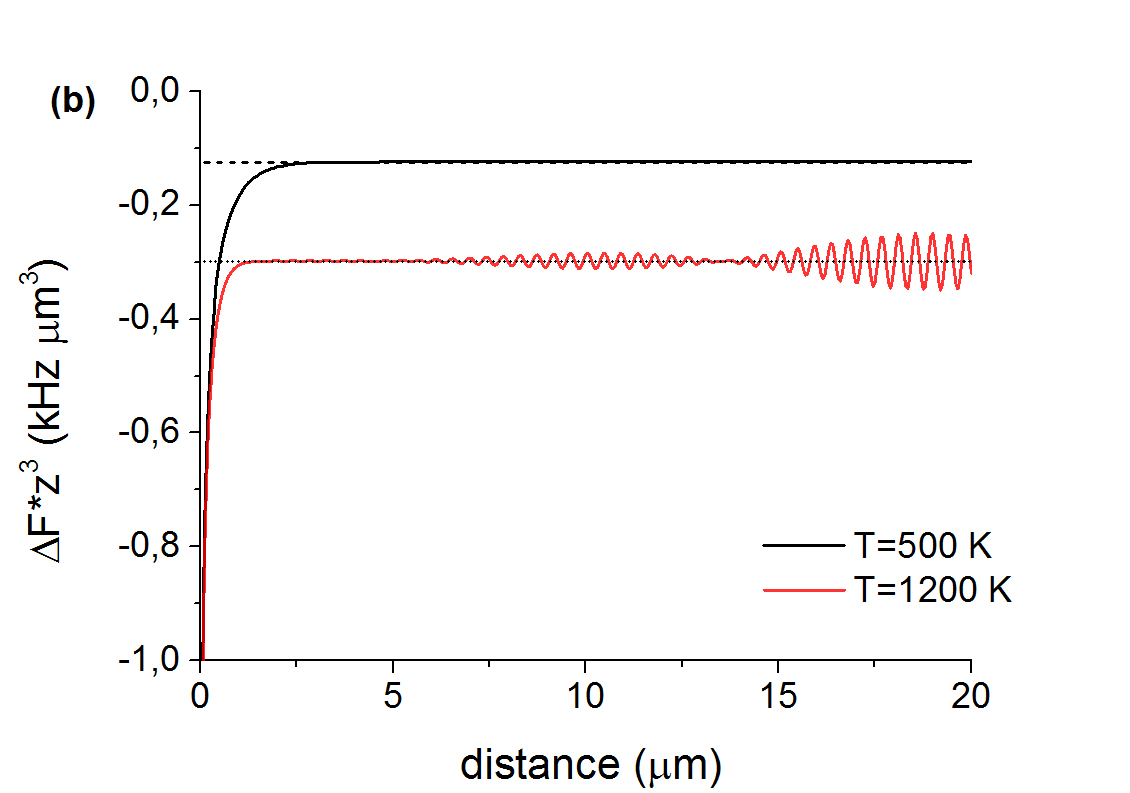}%
\caption{(Color online)(a) The CP free energy shift multiplied by $z^3$ for $\textrm{Cs}(6\textrm{S}_{1/2})$, $\textrm{Cs}^*(6\textrm{P}_{1/2})$ against a sapphire surface and the difference between them at two different temperatures, $T=500 K$ (black) and $T=1200 K$ (red or dark grey) for distances between 0 and 200 nm. (b) The CP free energy shift multiplied by $z^3$ for $\textrm{Cs}(6\textrm{S}_{1/2})$ against a sapphire surface for a distance range between 0 and 20$\mu$m, at two different temperatures, $T=500 K$ (black) and $T=1200 K$ (red or dark grey). The dashed and dotted lines show the results of the far field approximation of eqn. (16). For higher temperatures eqn. (16) is valid for distances closer to the surface.   \label{Fig5}}
\end{figure}

Finally, we perform a complete calculation of the Casimir-Polder free energy shift for the case of ground state Cs($6\textrm{S}_{1/2}$) and the low lying excited state $\textrm{Cs}^*(6\textrm{P}_{1/2}$). Here we take into account only the most important dipole couplings (Table II), ignoring the contribution of core excitations \cite{dereviankoPRL1999}. In Fig.\ref{Fig5} we show our results, focusing first (Fig.\ref{Fig5}a) on rather short distances from the sapphire wall. There is a dependence of the CP interaction on temperature, which is more important for the $\textrm{Cs}^*(6\textrm{P}_{1/2}$) state because the $ 6P_{1/2} \rightarrow 5D_{3/2} $ coupling is closer to the sapphire polariton frequency. The effect should be barely measurable with the precision of spectroscopic atom-surface interaction \cite{laliotisAPB2008} experiments. One notices that for the ground state Cs atom, the  van der Waals limit is not yet reached even at such short distances, characteristic of the distance dependence of the non-resonant contribution to the CP interaction \cite{WSpra1985, hindspra1991, gorzaEPJD2006, landraginpr1996}. Conclusive experimental evidence of this behaviour are only given for distances greater than 150 nm  \cite{benderprl2010}. This is less true for the $\textrm{Cs}^*(6\textrm{P}_{1/2}$) which seems to follow a $z^{-3}$ dependence for distances smaller than $ 100 \textrm{nm}$. Intuitively one can assume that this is because the wavelength of the dominant coupling is larger for  the $6\textrm{P}_{1/2}$ state (3$\mu \textrm{m}$ instead of 0.852$\mu \textrm{m}$ for $6\textrm{S}_{1/2}$ ), but this is not true. Remarkably it is the interplay between the different couplings (some virtual emissions and some virtual absorptions) that leads to this phenomenological adherence to the vdW $z^{-3}$ law. This coincidence could have important implications for spectroscopic experiments in the near field of the atom surface interaction, which are sensitive to the free energy \textit{difference}  between levels. For low lying excited states (such as the $6\textrm{P}_{1/2}$) experimental measurements of the CP shift could vary depending on the typical probing distance. In this particular example the $6\textrm{S}_{1/2} \rightarrow 6\textrm{P}_{1/2}$ shift is almost doubled between 0 and 100 nm and is clearly not following the vdW $z^{-3}$ law. This should be taken into consideration for experiments on low lying excited states of alkalis \cite{laliotisAPB2008, whitprl2014, commentbloch} (see also Discussion and Prospects section of \cite{laliotisSPIE2007}). It's worth mentioning that in the selective reflection experiments reported in \cite{laliotisAPB2008} the measured $C_{3}$ coefficient is 1.4 kHz $\mu \textrm{m}^{3} $, a value which is more consistent with Fig. \ref{Fig5} (for distances greater than 50-100 nm) than the theoretical vdW prediction of 0.8 kHz $\mu \textrm{m}^{3} $ (essentially for z=0).  Conversely high lying excited states exhibit huge CP shifts compared to low lying states and should be less affected by this problem \cite{chevrollieroptlett1991,laliotisnatcommun2014,passeratlaserphysics2014, failacheEPJD2003, failachePRL1999}. In Fig. \ref{Fig5}b we plot the CP shift of the Cs ground state atom at a much larger range of distances (0-20$\mu m$). The non-resonant term dominates and  the approximation of eqn. \ref{eqn16} is mostly valid in the far field. At very high temperatures (T=1200 K) one begins to see the beat between the QED oscillations of the resonant term at 894nm and 852nm. Putting these predictions to a real experimental test is at the moment extremely challenging.  

\section{Conclusions}

We have analysed the thermal effects of the CP interaction when the surface is at thermal equilibrium with the surrounding environment. We derived simple analytical expressions in the case of a dielectric with one surface resonance that provide a transparent physical interpretation of the temperature dependence of the CP free energy shift of atoms or molecules. Our work shows that in the near field thermal effects are entirely due to the excitation of evanescent surface modes (near field thermal emission). The results can be easily extended to the case of metals where plasmon frequencies are typically at UV wavelengths. Using more realistic models for dielectrics is also straightforward but significantly more tedious. We also derive simple expressions in the far field approximation, valid not only for ground state but also excited state atoms and molecules. We show that retardation effects can be significant for spectroscopic experiments on low lying states performed at nanometric distances away from the surface. This raises important questions concerning the validity of the $z^{-3}$ vdW law. It is often neglected that the CP interaction can also affect the radiative properties (transition rates) of atoms or molecules \cite{HenkelAPB1999, BuhmannPRA2008}. Our conclusions can be extended to calculate a distance dependent transition linewidth \cite{WSpra1984,WSpra1985} using the imaginary part of the linear susceptibility matrix. Although our treatment was entirely performed in the case of thermal equilibrium, our demonstration that the near field temperature dependence is solely due to the excitation of evanescent surface waves suggests that in the near field our results are also valid in an out of equilibrium case with T being the temperature of the surface \cite{DorofeyevLP2013}. A complete treatment of the out of equilibrium case for excited state atoms or molecules is, however, more challenging. Finally, the simplicity of our results could render them very useful when applied to intrinsically more complicated problems such as multi-layered dielectrics \cite{WSpra1985}, or 2-D dielectrics such as graphene deposited on dielectric substrates \cite{mostepanenkopra2014} or even meta-surfaces. 

\begin{acknowledgments}
We would like to thank M-P Gorza for help with the theoretical evaluation of the linear susceptibility and discussions on the effects of retardation. We would also like to thank D. Bloch for discussions and for providing comments on the manuscript. We finally thank T. Passerat de Silans and I. Maurin for useful conversation.
\end{acknowledgments}

\bibliography{biblio}

\end{document}